\documentclass{iau}
\usepackage{graphicx}
\usepackage{natbib}

\title[The Debiased Kuiper Belt] 
{The Debiased Kuiper Belt: Our Solar System as a Debris Disk}

\author[S.~Lawler et al.]   
{Samantha M. Lawler$^1$
 \and the CFEPS Team}

\affiliation{$^1$Department of Physics and Astronomy \\ 
University of British Columbia \\
6224 Agricultural Road \\ 
Vancouver, BC V6T 1Z1, Canada \\
email: {\tt lawler@astro.ubc.ca} \\[\affilskip]}

\pubyear{2013}
\volume{299}  
\pagerange{232-236}
\jname{Exploring the Formation and Evolution of Planetary Systems}
\editors{M.~Booth, B.~C.~Matthews \& J.~R.~Graham, eds.}
\begin{document}

\maketitle

\begin{abstract}
The dust measured in debris disks traces the position of planetesimal belts. In our Solar System, we are also able
to measure the largest planetesimals directly and can extrapolate down to make an estimate of the dust. The
zodiacal dust from the asteroid belt is better constrained than the only rudimentary measurements of Kuiper belt
dust. Dust models will thus be based on the current orbital distribution of the larger bodies which provide the
collisional source. The orbital distribution of many Kuiper belt objects is strongly affected by dynamical interactions
with Neptune, and the structure cannot be understood without taking this into account.
We present the debiased
Kuiper belt as measured by the Canada-France Ecliptic Plane Survey (CFEPS). This model includes the absolute
populations for objects with diameters $>$100~km, measured orbital distributions, and size distributions of the
components of the Kuiper belt: the classical belt (hot, stirred, and kernel components), the scattering disk, the
detached objects, and the resonant objects (1:1, 5:4, 4:3, 3:2 including Kozai subcomponent, 5:3, 7:4, 2:1, 7:3,
5:2, 3:1, and 5:1).
Because a large fraction of known debris disks are consistent with dust at Kuiper belt distances
from the host stars, the CFEPS Kuiper belt model provides an excellent starting point for a debris disk model, as
the dynamical interactions with planets interior to the disk are well-understood and can be precisely modelled using
orbital integrations.

\end{abstract}

\firstsection 
\section{Introduction}

There is much discussion in the literature about using resonant structure in debris disks as an indirect way to 
find planets \citep[e.g.][]{Wyatt2006,StarkKuchner2008}.
In debris disks, the observable is dust grains, which is used to infer the presence of a 
collisionally-grinding planetesimal belt \citep{Wyatt2008}.
In our own Solar System, the opposite is true: we can directly observe orbits of the planets and planetesimals to great precision, 
while no direct observations exist of dust in the Kuiper belt.
We also know that the Kuiper belt has significant resonant structure \citep{Gladmanetal2012},
which requires that Neptune's orbit has migrated outwards. 
This migration may have happened smoothly as Neptune scattered planetesimals inwards onto Jupiter-crossing orbits,
``snowplowing'' objects into mean-motion resonances along the way
\citep{ChiangJordan2002,HahnMalhotra2005}, or Neptune may have been scattered outward by a close
encounter with one of the other giant planets, and trapped planetesimals in mean-motion resonances as
its eccentric orbit damped to near-circular \citep[the ``Nice model'';][]{Levisonetal2008}.

\section{Survey Strategy}

The Canada-France Ecliptic Plane Survey (CFEPS) was a large survey carried out with the Megacam instrument on the 
Canada-France-Hawaii Telescope (CFHT) between 2003--2007, with the goal of discovering many trans-Neptunian objects (TNOs)
and using the known sky coverage, obervation depths, and tracking fractions 
in different observing blocks at many points along the ecliptic plane to debias
the observations and regain the true orbital distribution of the Kuiper belt.

The observing strategy for CFEPS is described in detail in \citet{Jonesetal2006}, 
\citet{Kavelaarsetal2009} and \citet{Petitetal2011}.
To summarize, a patch of sky is observed three times in a night, with each observation separated by
about one hour.  
These triplets of images are searched for moving objects, and preliminary orbits are calculated.
One month later, followup images are taken of the same field, and the same objects are recovered.
This is repeated each month for 4--5 months.  
This 4--5 month baseline is used to predict object positions one year later, and the objects are recovered
again, further improving the calculated orbits.

A 10~Myr-long orbital integration is performed using the orbital elements and position of each TNO, 
and including the gravitational effects of the four giant planets.
The TNOs are then placed into dynamical classes following the procedure outlined by 
\citet{Gladmanetal2008}.
The classes are resonant (TNOs in mean-motion resonances with Neptune),
scattering (TNOs that closely encounter Neptune within 10~Myr), 
detached (TNOs that currently do not dynamically interact with Neptune),
and classical (TNOs on low-$e$, low-$i$ orbits with $a$ within a few AU of Neptune).

\section{Discoveries and Debiasing}

\begin{figure}
\begin{center}
\includegraphics[scale=0.26]{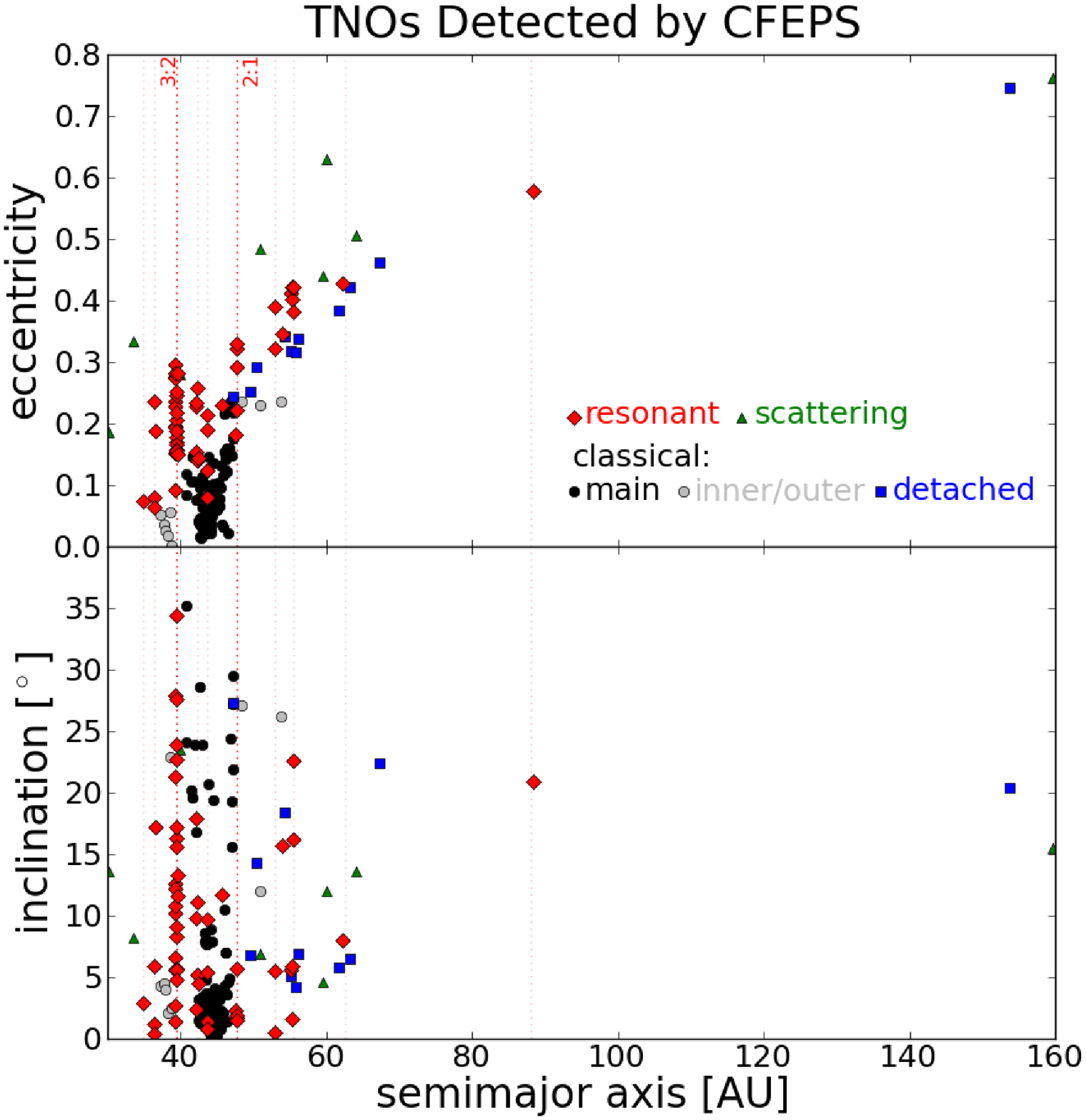} \includegraphics[scale=0.26]{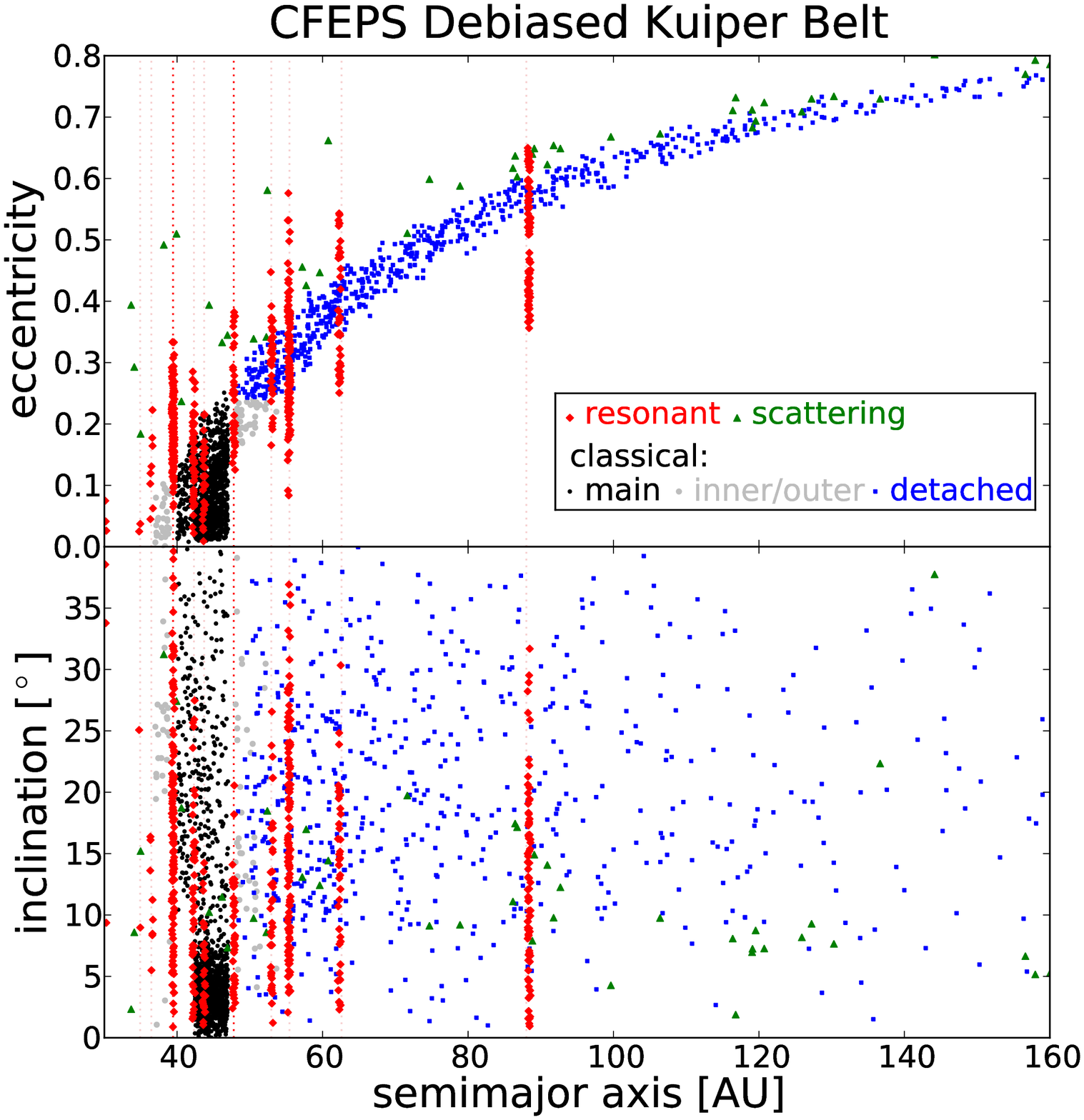} 
 \caption{Left panel shows orbital elements ($a$,$e$,$i$) of TNOs discovered by CFEPS, 
and right panel shows debiased TNO orbital element distributions,
where the relative numbers in each population and in each resonance are correct.
Objects have symbols according to dynamical class (see key), and mean-motion resonance positions are indicated
by dotted lines.}
   \label{debiasing}
\end{center}
\end{figure}

CFEPS detected 176 TNOs.  
Their orbital element distributions are shown in the left panel of Figure~\ref{debiasing}.

\begin{figure}
\begin{center}
\includegraphics[scale=0.26]{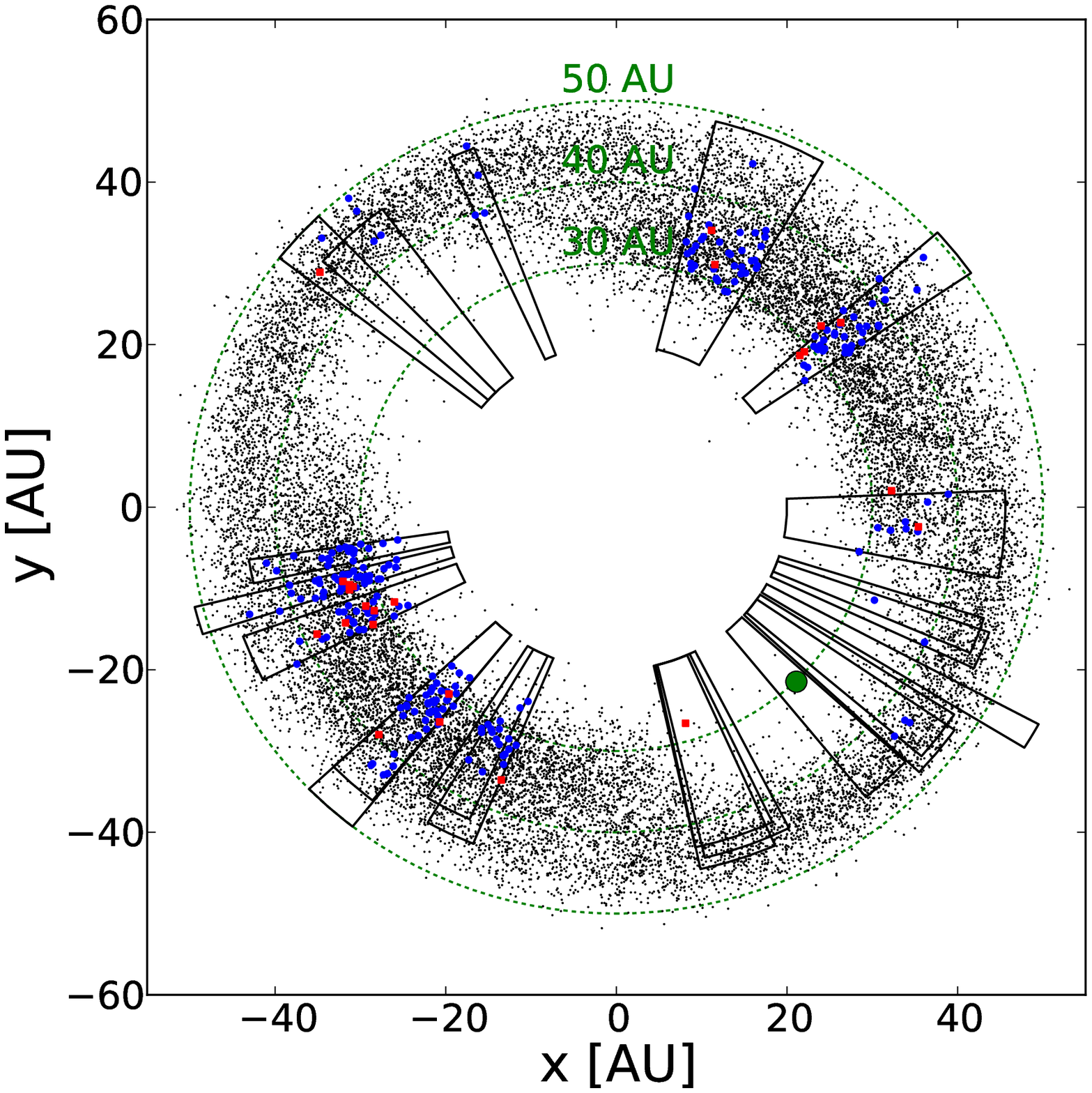} \includegraphics[scale=0.34]{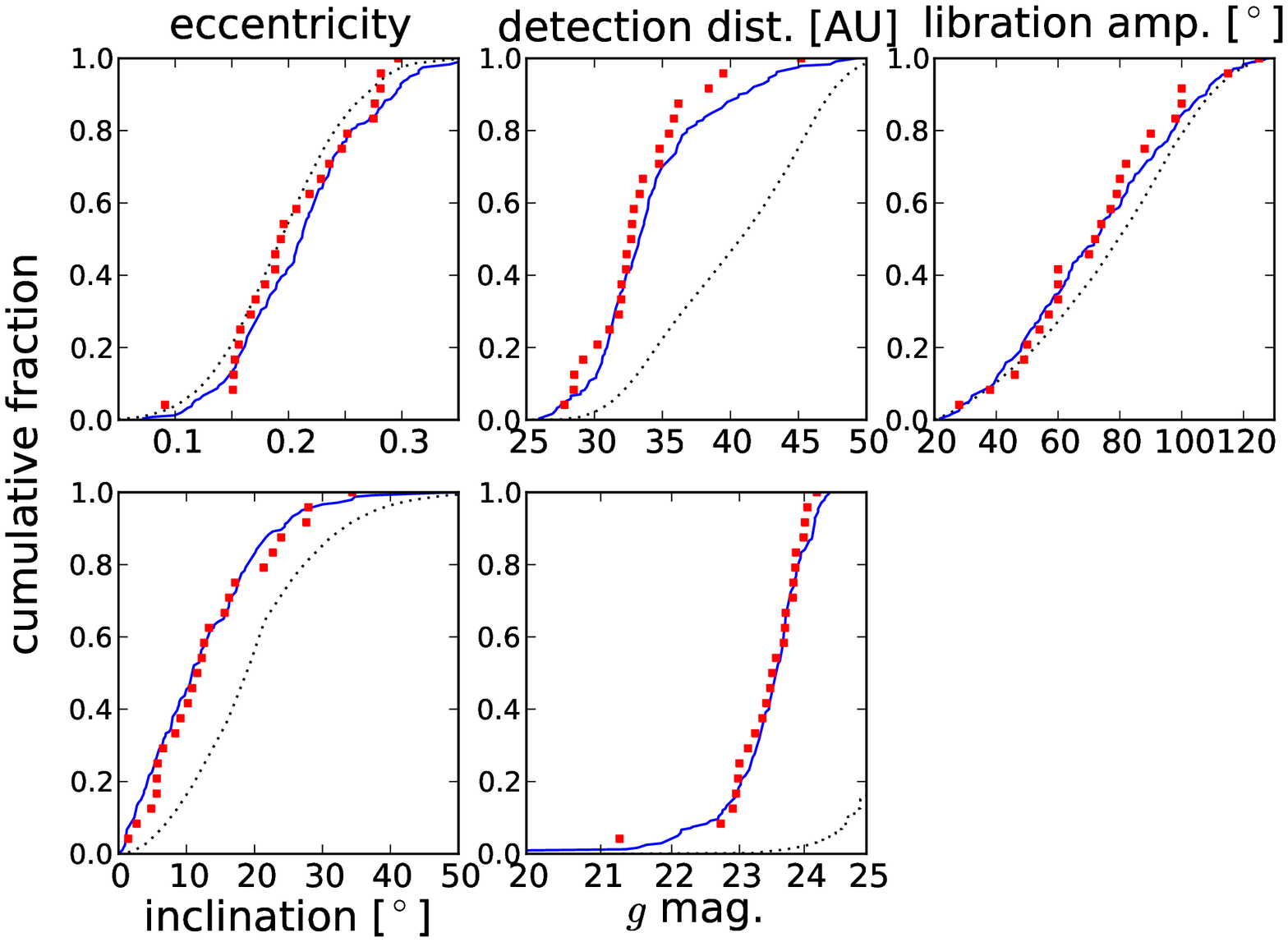} 
 \caption{Left panel shows a distribution of plutinos viewed from above the Solar System, 
and right panel shows cumulative distributions in five different observable parameters: 
eccentricity, distance at detection, libration amplitude, inclination, and visible $g$ magnitude.
Black (in both panels) shows synthetic non-detections, blue circles/solid lines show synthetic detections by the 
survey simulator, and red squares shows real detections.
Also shown in the left panel are the CFEPS observing blocks, with the relative magnitude limits indicated 
by the outer edge of each wedge (showing the distance limit where a $\sim$100~km diameter object would still be detected).}
   \label{cumuplots}
\end{center}
\end{figure}

Because the survey is well-characterized, it can be debiased to reproduce not only the 
orbital elements of each different dynamical class, but also the absolute populations.
We will use the 
plutinos (objects in the 3:2 mean-motion resonance) as an example 
to explain the debiasing process (Figure~\ref{cumuplots}).

First, we choose parameters to model the distribution of synthetic plutinos
in eccentricity, inclination, libration amplitude, and size distribution (shown in black in Figure~\ref{cumuplots}),
which is discussed in detail in \citet{Gladmanetal2012} and \citet{LawlerGladman2013}.
This synthetic distribution is run through the CFEPS survey simulator, where each object is evaluated 
to determine if it would have been detected by CFEPS.
Synthetic detections are shown in blue in Figure~\ref{cumuplots}.
The cumulative distributions of five parameters of the synthetic detections are compared statistically
to the cumulative distributions of the real CFEPS detections.
Any modeled distribution where the bootstrapped comparison statistic
for any of the five distributions occurs $<$5\% of the time is considered rejected.
In this way, we get very powerful constraints on the orbital element distributions and 
the size distribution, and we can use this to calculate absolute population.

This modelling process is repeated for each resonance \citep{Gladmanetal2012} and other Kuiper belt components individually
\citep{Petitetal2011}.

\section{A Simple Kuiper Belt Dust Model}

Many very detailed models of the dust produced by the Kuiper belt exist in the literature 
\citep[e.g.][]{Vitenseetal2010,KuchnerStark2010}.
The model presented here is only meant to serve as a demonstration of how
the CFEPS debiased Kuiper belt distribution (available for download at \texttt{www.cfeps.net})
can be used as a starting point for a Kuiper belt dust simulation.

\begin{figure}
\begin{center}
\includegraphics[scale=0.25]{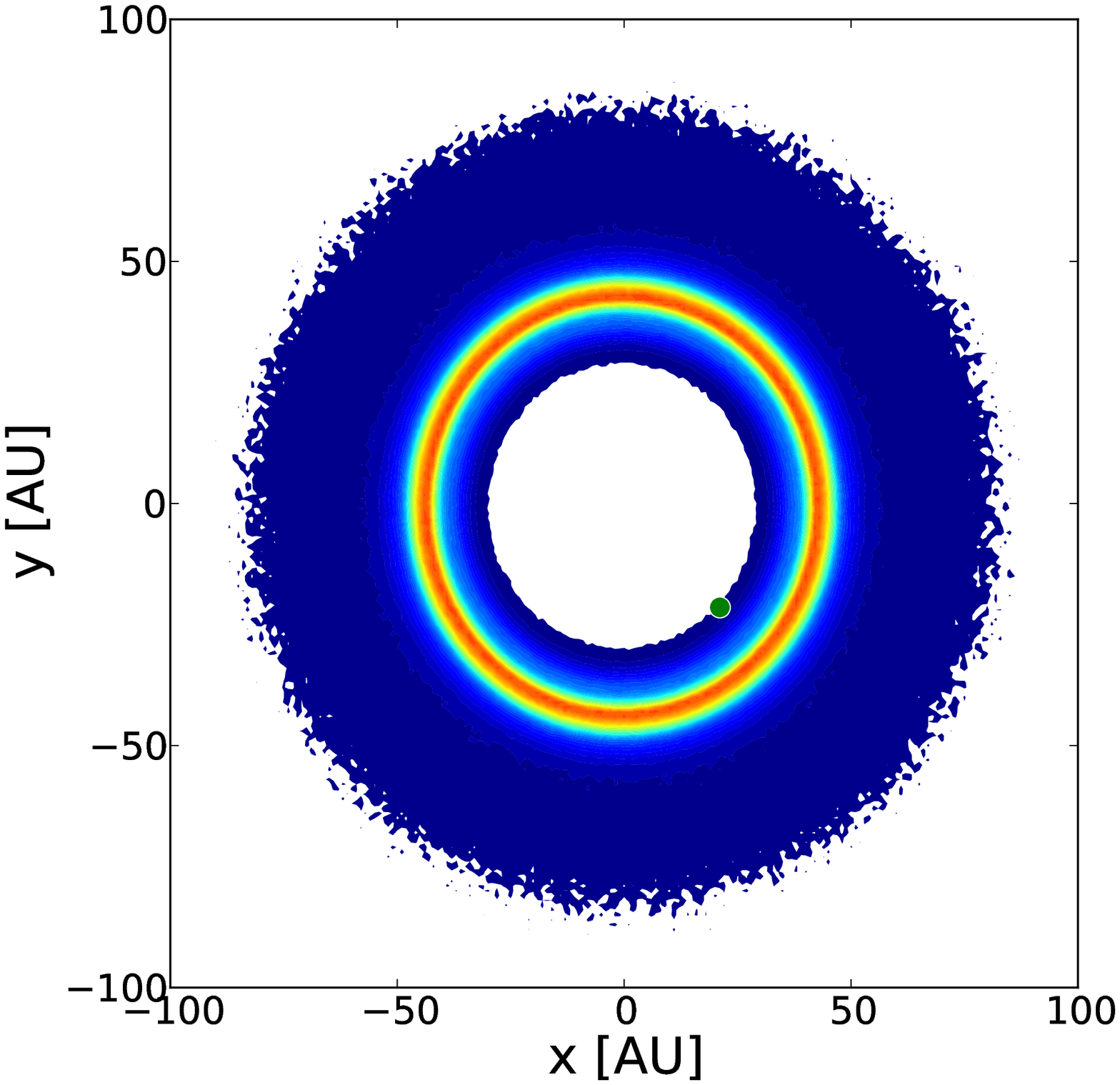} \includegraphics[scale=0.25]{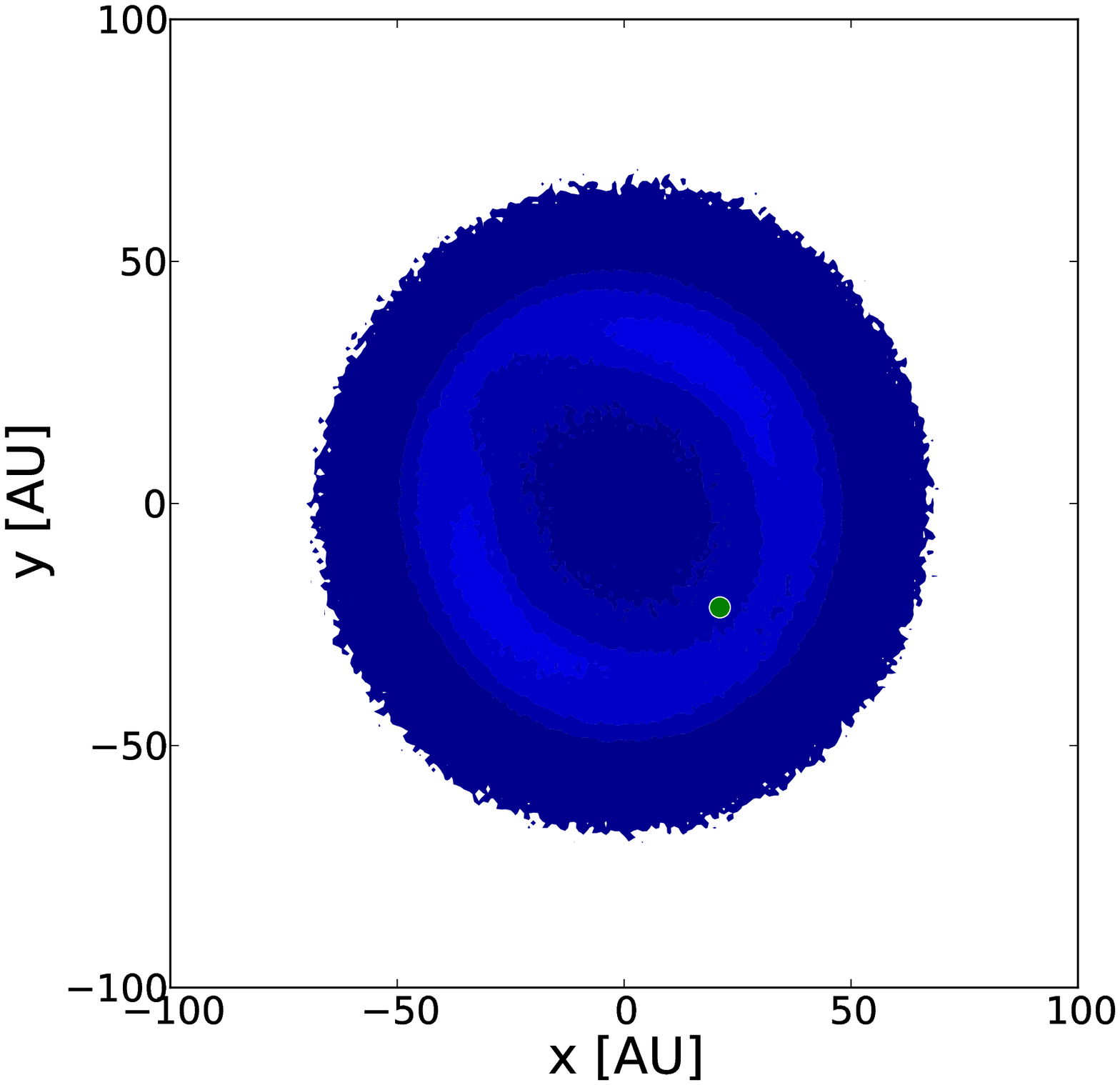} 
\includegraphics[scale=0.25]{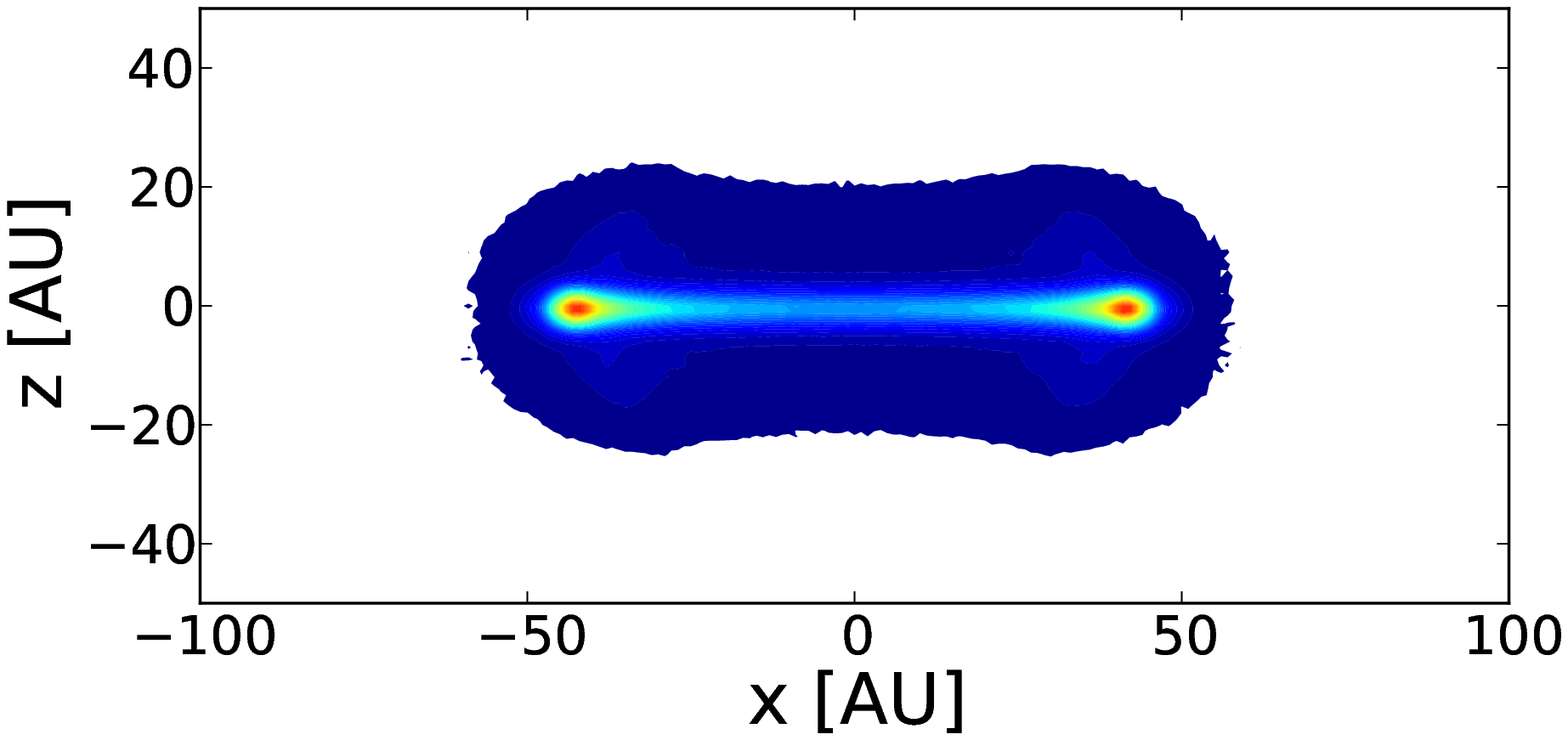} \includegraphics[scale=0.25]{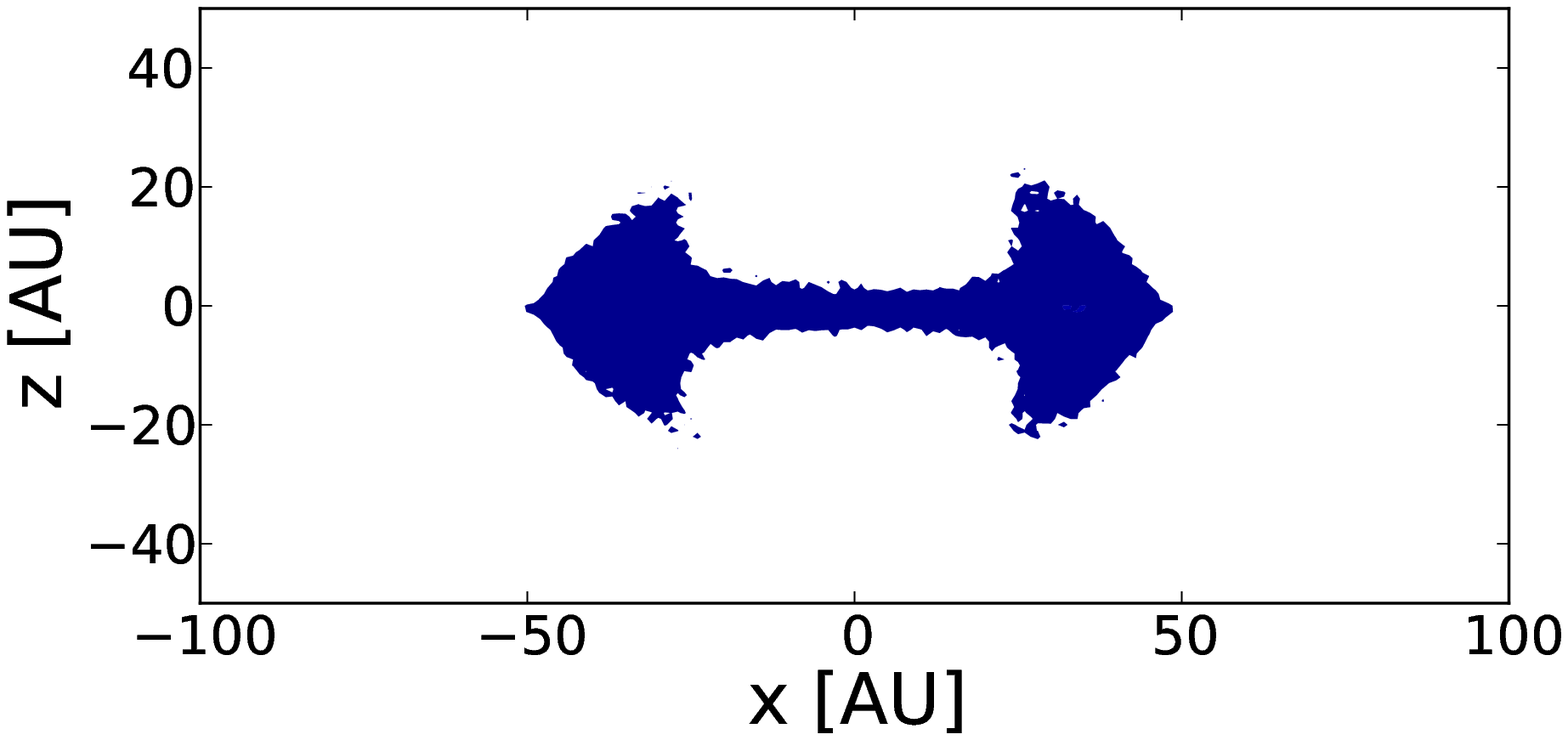}
 \caption{Number density contour plots of the classical (left panels) and resonant (right panels) debiased 
Kuiper belt looking down on the Solar System from above (top panels) and from edge-on (bottom panels).  
The position of Neptune is shown by a green circle in the top panels.
The two top panels have the same contour levels, and the two bottom panels have the same contour levels, making
it obvious that the classical objects dominate the number density.}
   \label{nums}
\end{center}
\end{figure}

Figure~\ref{nums} shows contour plots of the number density of the debiased resonant (right panels) 
and non-resonant TNOs (left panels).
Applying a simple dust flux model, where each TNO is replaced by a modified-blackbody dust grain,
the resonant structure is completely drowned out by the much higher number density of classical objects.

\section{Conclusions}

The Kuiper belt has been sculpted by giant planet migration, and contains many populated resonances
(20\% of $>$100~km objects are resonant), but this is difficult, if not impossible, 
to see in the overall disk structure as seen from afar.
Currently, no theoretical giant planet migration models can reproduce all the structure that is observed
in the Kuiper belt \citep{Gladmanetal2012}.
Because of the completely unknown biases that are present in the Minor Planet Center Database,
the CFEPS debiased Kuiper belt distribution is currently the most accurate model for TNOs $>$100~km in diameter,
and makes an excellent starting point for collisional dynamical Kuiper belt dust models.

\end{document}